\documentclass{JHEP3}
\usepackage{epsfig}
\usepackage{graphicx}
\usepackage{dcolumn}
\usepackage{bm}
\newcommand{\be}{\begin{equation}}
\newcommand{\ee}{\end{equation}}
\def\bea{\begin{eqnarray}}
\def\eea{\end{eqnarray}}

 \newcommand{\ft}[2]{{\textstyle\frac{#1}{#2}}}
 \newcommand{\Ka}{K{\"a}hler}

 \def\be{\begin{equation}}
\def\ee{\end{equation}}
\def\bea{\begin{eqnarray}}
\def\eea{\end{eqnarray}}

\def\lesssim{\mathrel{\hbox{\rlap{\hbox{\lower4pt\hbox{$\sim$}}}\hbox{$<$}}}}
\def\gtrsim{\mathrel{\hbox{\rlap{\hbox{\lower4pt\hbox{$\sim$}}}\hbox{$>$}}}}
\title{Supersymmetry and Stability of  Flux Vacua}

\author{Jose J. Blanco-Pillado$^{1}$, Renata Kallosh$^{2}$ and Andrei Linde$^{2}$
\\
${}^{1}$Center for Cosmology and Particle Physics, Department of Physics, New~York~University, 4 Washington Place, New York, NY 10003\\
   ${}^{2}$Department of Physics, Stanford University, Stanford, CA 94305,
USA,\\ and 
Kyoto University, Yukawa Institute, Kyoto, 606-8502, Japan }
\received{\today}       
 \preprint{YITP-05-49  \\
 November 3, 2005}

\abstract{
We describe a modified KKLT mechanism of moduli stabilization in a supersymmetric Minkowski vacuum state. In this mechanism, supersymmetry ensures vacuum stability and positivity of the mass matrix for the dilaton, complex structure, and  volume modulus.}

\begin{document}

\section{Introduction.}

In the last few years, there has been a significant progress in solving the problem of moduli stabilization in string theory. 
We should mention, in particular, a mechanism for stabilizing the dilaton and the complex structure moduli by fluxes 
\cite{Giddings:2001yu} and the KKLT construction, which allows, in principle, to fix all remaining moduli, including the 
volume modulus describing the size of the compactified space  \cite{Kachru:2003aw}.

The standard way towards moduli stabilization in the KKLT construction consisted of two parts. One should first fix the 
axion-dilaton and the complex structure moduli by turning on fluxes following  \cite{Giddings:2001yu}. In order to do this,
one should find a minimum of the potential in the effective N=1 supergravity theory with a superpotential given by
\begin{equation}\label{flux}
  W_{\rm flux} = \int G_3\wedge \Omega= f\cdot \Pi(x) -\tau h\cdot \Pi(x) \ .
\label{super1}
\end{equation}
Here $f$ is a flux vector from the RR 3-form, $\tau$ is the axion-dilaton, $h$ is a  flux vector from the NS form and $\Pi $ 
is a period vector depending on complex structure moduli $x$. 

Once the moduli $\tau$ and $x$ are fixed at some values $\tau_{0}$ and $x_{0}$,  one should fix the volume modulus by adding
to the superpotential a term $W_{\rm np}(\rho)$ describing nonperturbative effects (e.g. gluino condensation). The simplest 
nonperturbative term of this type can be written as 
\be\label{oneexp}
W_{\rm np}= Ae^{-a\rho} \ .
\label{adssup1}
\ee

In practice, the procedure of finding a minimum of the scalar potential $V(\tau,x,\rho)$ was replaced by a simpler procedure 
of finding its supersymmetric extremum and making sure that this extremum is a minimum with respect to the volume modulus. 

However, this is not the end of the story: The supersymmetric extremum of the potential in N=1 supergravity usually occurs 
at negative values of the potential (AdS extremum). One must then uplift this minimum to make the value of the potential slightly positive
at that point, so that $V_{0} \sim 10^{{-120}}$ in Planck units. The uplifting is achieved by adding to the potential a term $C\over \rho^{2}$ corresponding to the energy of the $\overline {D3}$ brane \cite{Kachru:2003aw,Kachru:2003sx} or $D7$ brane \cite{Burgess:2003ic}. Some models of flux stabilization lead directly to dS vacua \cite{Saltman:2004sn}.

Properties of the potential after stabilization of the axion-dilaton and complex structure moduli were studied in \cite{Denef:2004ze}. The authors found that  a certain fraction of all AdS extrema corresponds to saddle points. This is not a problem in AdS because a supersymmetric extremum in AdS can be stable even if it is a maximum or a saddle point \cite{Breitenlohner:1982bm}. However,  after the uplifting, AdS space becomes dS space, and the fields near a maximum or a saddle point of the scalar potential  become unstable.

This is not a real problem since a considerable fraction of all AdS vacua studied in \cite{Denef:2004ze} are in fact minima. However, the problem may reappear again when one tries to stabilize all fields including the volume modulus: Even if the dilaton and volume modulus may seem stabilized, one may encounter instabilities with respect to their simultaneous evolution \cite{Choi:2004sx,deAlwis:2005tf}. The root of the problem is the same: One should carefully check whether the AdS extremum prior to the uplifting is a minimum but not a saddle point with respect to all moduli.

Another set of problems appears when one tries to analyze a possible impact of the uplifting term $C\over \rho^{2}$ on the properties of the extremum. A nice property of the original KKLT model was that the nonperturbative potential induced by the superpotential (\ref{oneexp}) was very sharp, so an addition of the slowly decreasing function  $C\over \rho^{2}$ did not destabilize the minimum. However, it changed the position of the field $\rho$. For some other nonperturbative potentials, uplifting may lead to destabilization, either with respect to the volume modulus or with respect to some other moduli. So in general one should make an additional investigation to check that the extremum of the effective potential remains a minimum after the uplifting.

Over the last few years many  examples of all moduli stabilization in type IIB theory have been constructed \cite{Denef:2005mm}. Many of them  show that in fairly generic examples of the KKLT type, one can find  explicit  class of models with de Sitter vacua  which are tachyon free and the potential is a local minimum in all directions.  These examples support the general statistical analysis in \cite{Denef:2004ze} which argues that there are many examples without tachyons. However, one does have to check each new model and verify that it is tachyon-free with respect to all moduli. 

The main idea of our paper is to stabilize all moduli in a supersymmetric Minkowski vacuum, extending the approach of Ref.  \cite{Kallosh:2004yh}. As we will show, the absence of tachyons in such models is guaranteed by construction.

This issue is closely related to the problem of supersymmetry breaking in the KKLT scenario. Different aspects of this problem were studied by many authors; see e.g.  \cite{Susskind:2004uv} and references therein. One feature of supersymmetry breaking is particularly important in the cosmological context. In the original version of the KKLT scenario, the height of the barrier separating the minimum of the potential from the 10D Minkowski vacuum is $O(m^{2}_{3/2})$. If one tries to achieve inflation by adding to the KKLT potential a contribution of the inflaton field, it always comes divided by a certain degree of $\rho$. Therefore adding a large inflaton contribution may destabilize the volume modulus. One can show that the barrier disappears and the universe decompactifies if the energy density of the inflaton field (or of any other fields living in 4D) becomes much greater than the height of the barrier.  This implies that the KKLT minimum appears only very late in the cosmological evolution, when the Hubble constant becomes smaller than the gravitino mass  \cite{Kallosh:2004yh}.

There are some ways to slow down the rolling of the volume modulus at the first stages of the evolution of the universe and avoid decompactification, see e.g. \cite{Berndsen:2005qq,Linde:2005dd}. However, for many years it was believed that the gravitino mass is extremely small, $m_{3/2} =O(1) \rm{TeV} \sim 10^{-15} M_{p}$. In this case one would come to an unexpected conclusion that the   vacuum stabilization in string theory is a very low energy scale phenomenon. This would also mean that the Hubble constant during inflation cannot be greater than $m_{3/2} \sim 10^{-15} M_{p}$. Whereas such low-scale inflationary models are possible, they are highly unconventional and require introduction of several extremely small parameters.

There are two ways to alleviate this problem. First of all, one may try to construct the models with $m_{3/2} \gg 10^{-15} M_{p}$, see e.g. \cite{DeWolfe:2002nn,Arkani-Hamed:2004fb,Balasubramanian:2005zx}. Another approach is to modify the original KKLT mechanism  \cite{Saltman:2004sn,Kallosh:2004yh,Balasubramanian:2005zx}. In this paper we will study this possibility following Ref. \cite{Kallosh:2004yh}. It was shown there that if one considers racetrack superpotentials with two exponents, $W_{\rm np}= Ae^{-a\rho}+ Be^{-b\rho} $, one can find such parameters that the scalar potential has a supersymmetric minimum at $V_{0}=0$, i.e. in Minkowski space, even before an uplifting.  In this approximation, the gravitino mass vanishes. By allowing a small  supersymmetry breaking and using a very small uplifting, one can make the height of the barrier stabilizing the KKLT minimum, as well as the mass squared of the volume modulus, many orders of magnitude  greater than $m^{2}_{{3/2}}$. This allows inflation at $H \gg m_{3/2}$.

In the present paper we will show that the same mechanism allows to alleviate the vacuum instability problems  discussed above. The reason is very simple: If, following \cite{Kallosh:2004yh}, we find a local supersymmetric extremum of the scalar potential {\it in Minkowski space}, then the corresponding extremum will be stable in Minkowski space due to the positive energy theorem in supergravity \cite{Deser:1977hu}. We will demonstrate this explicitly by the investigation of  the properties of the effective potential near the supersymmetric extremum; for a similar approach see also \cite{Ferrara:1994kg}. 

Of course, we still need to break supersymmetry and slightly uplift the potential to $V_{0} \sim 10^{{-120}}$. However, this should not destabilize the potential if the gravitino mass is much smaller than the masses of the stabilized moduli. 

\section{Convexity of the F-term potential in supersymmetric Minkowski vacua.}

It will be useful for our purposes to present here the general compact expression for the second derivative of the F-term potential
in N=1 supergravity in terms of the fermion masses in the supersymmetric vacuum. We will consider the most general case of an arbitrary 
number of fields $z_{a}$, arbitrary K\"ahler potential ${\cal K}(z_{a}, \bar z_{\bar a})$ and arbitrary holomorphic superpotential $W(z_{a})$.  

We use the notation close to \cite{Kallosh:2000ve} where the  N=1 supergravity action with fermionic terms is presented in a form useful 
for the study of the vacua of the effective theory. We use units in which $M_P=1$. In addition to the holomorphic superpotential $W(z_{a})$,
we define a covariantly holomorphic complex gravitino mass as, 
\begin{equation}
e ^{ K/2}W \equiv {\it m}(z_{a}, \bar z_{\bar a})\ , 
\label{defm}
\end{equation}
which is related to the (real) gravitino mass,
\begin{equation}
  M^2_{3/2}  =|m \, \bar m| \ .
\label{defmee}
\end{equation}
The complex gravitino mass is covariantly holomorphic, which means that 
\begin{equation}
  \bar D_{\bar a}m\equiv (\partial_{\bar a} -{1\over 2} K_{\bar a} ){\it m}=0\ , \qquad D_{ a}\bar m\equiv (\partial_{ a} -{1\over 2} K_{ a} )\bar {\it m}=0 \ .
\label{cov}
\end{equation}
 We also define the \Ka\ covariant derivatives as,
\begin{eqnarray}
D_a {\it m} =  \, \partial _a {\it m}
+\ft12 (\partial _a { K}) {\it m}\equiv {\it m}_a\ , \qquad \bar D_{\bar a} \bar {\it m} =  \, \partial _{\bar a}\bar {\it m}
+\ft12 (\partial _{\bar a}{ K}) \bar {\it m}\equiv \bar {\it m}_{\bar a} \ .
\end{eqnarray}
The complex masses of the chiral fermions in N=1 supergravity are equal to
\begin{equation}
  D_a D_b {\it m} \equiv {\it m}_{ab}\ , \qquad \bar D_{\bar a}\bar D_{\bar b}\bar {\it m}\equiv \bar {\it m}_{\bar a  \bar b} \ .
\label{chiral}
\end{equation}
The \Ka\ metric is
$g_{a\bar b}  =\partial_a \partial_{ \bar b}{K}$.
Using this notation we can rewrite the standard F-term potential $V= e^{\cal K}( |DW|^2-3|W|^2)$ as follows
\begin{equation}\label{pot}
V={\it m}_a\, g^{a\bar b} \, \bar {\it m}_{\bar b}  -3|{\it m}|^2 \equiv |{\it m}_a|^2-3|{\it m}|^2 \ .
\end{equation}
The first derivative of the potential becomes,
\begin{equation}
\partial _a V=D_a V= -2 {\it m}_a\,\bar {\it m} +{\it m}_{ab}g^{b\bar b} \,\bar {\it m}_{\bar b} \ .
\label{diV}
\end{equation}
where we have taken into account that $ D_{ a} \bar {\it m}_{\bar b}= g_{ a \bar b} \bar {\it m}$ and that the potential is \Ka\ invariant.

From now on we focus only on supersymmetric vacua, i.e.,
\begin{eqnarray}
D_a {\it m}  \equiv {\it m}_a =0 \ , \qquad \bar D_{\bar a} \bar {\it m}  \equiv \bar {\it m}_{\bar a}=0\ ,  \qquad \Rightarrow \qquad \partial _a V=0 \ .
\end{eqnarray}
In this case, the second derivatives of the potential at $\partial _a V=0$ are given by
\begin{equation}
  \partial_{b} \partial _a V =-{\it m}_{ba} \bar {\it m}\ , \qquad \partial_{\bar b} \partial _{\bar a} V =-\bar {\it m}_{\bar b \bar a} {\it m} \ ,
\label{holom}
\end{equation}
\begin{equation}
  \partial_{\bar b} \partial _a V= -2 g_{\bar b a} {\it m} \bar {\it m} + {\it m}_{ac}g^{c\bar c} \,\bar {\it m}_{\bar c \bar b} \ .
\label{mix}
\end{equation}
In general we see that the AdS critical point is a saddle point since there are positive and negative directions at ${\it m}\neq 0$. 
That is why uplifting of the supersymmetric AdS extremum may lead to an instability. This instability is not unavoidable though. 
In particular, it disappears if the complex gravitino mass  $\bar {\it m}$ is much smaller than all moduli masses.

This fact becomes manifest when the minimum of the potential occurs at vanishing value of the cosmological constant. 
In this case, one has ${\it m}= 0$ and ${\it m}_a=0$ at the supersymmetric critical point,  
so we can find the following matrix $\cal M$,
\begin{equation}
 {\cal M}_{ba} = \partial_{b} \partial _a V =0\ , \qquad {\cal M}_{\bar b \bar a}=\partial_{\bar b} \partial _{\bar a} V =0  \ ,
\label{holom1}
\end{equation}
\begin{equation}
{\cal M}_{a \bar b} = \partial_a \partial_{\bar b} V=  {\it m}_{ac}g^{c\bar c} \,\bar {\it m}_{\bar c \bar b} \ . 
\label{mix1}
\end{equation}
This is a positive definite matrix, and consequently 
all its eigenvalues are positive.  Indeed, for an arbitrary
vector $\phi$,
\begin{equation}
\phi_a {\cal M}_{a \bar{b}}{\bar \phi_{\bar b}}= \phi_{a} {\it m}_{ac} \, g^{c\bar c} \,\bar {\it m}_{\bar c \bar b} \,
 \bar\phi_{\bar b} = \Phi_{c}\, g^{c\bar c} \,\bar \Phi_{\bar c}\  , 
\end{equation}
where  $\Phi_{c} = \phi_{a} {\it m}_{ac}$. We now recall that the matrix $g^{c\bar c}$ is the inverse matrix of the \Ka\ metric
and therefore is required to be positive definite, so we have 
$
\Phi_{c}\, g^{c\bar c} \,\bar \Phi_{\bar c} \geq 0
$
for any vector $\Phi$. Thus  $\cal M$ is  a positive 
definite matrix, namely,
\begin{equation}
\phi_a \, {\cal M}_{a \bar{b}} \, {\bar \phi_{\bar b}} \geq 0 \ .
\end{equation}
This is true for an arbitrary vector $\phi_{a}$, which means that the second 
derivative of the potential is non-negative in all directions in the moduli space.

One can easily understand the meaning of this result in terms of equation (\ref{pot}): $V= |{\it m}_a|^2-3|{\it m}|^2$. 
In a supersymmetric extremum with $V = 0$, the function $m$ is of the second order in the deviation from this extremum 
$\delta z$, because both $m$ and its first derivative ${\it m}_a$ vanish at the extremum. Therefore the negative term 
$-3|{\it m}|^2$ is quartic with respect to $\delta z$;  the only term quadratic in $\delta z$ is the non-negative 
term $ |{\it m}_a|^2$.

One potential caveat in this discussion is the possibility that the mass matrix may vanish in some directions. This may occur, 
e.g., if the potential is quartic in $\delta z$. However, one can generalize the argument given above, expanding $m$ in all 
higher powers of $\delta z$, and show that   ${\it m}_a$ always contains terms lower order in $\delta z$ as compared with 
${\it m}$. Therefore at small $\delta z$ the positive term  $|{\it m}_a|^2$ dominates and the potential is convex. A similar argument can be found in \cite{Dine:1987vf}.

The only exception from this rule may occur if all higher order terms in expansion of ${\it m}$ in terms of $\delta z$ 
vanish in some direction, which corresponds to a flat direction of the potential. However, one often finds 
(as we will find in an example to be considered in Section \ref{example}), that the solutions of equations 
${\it m} = 0$ and ${\it m}_a = 0$  represent a discrete set of points instead of a line of supersymmetric points. 
In this case one can show that the potential cannot have flat directions, so all moduli are indeed stabilized. 
To prove it, one may assume that the potential has a flat direction along which  $V = |{\it m}_a|^2-3|{\it m}|^2= 0$. 
This condition can be considered as a differential equation on ${\it m}$ with the initial condition ${\it m} = 0$, ${\it m}_a=0$, 
for all $a$. One can easily check that the only solution of this equation along the flat direction is ${\it m} = 0$, ${\it m}_a=0$, 
which means that supersymmetry should take place along the valley. In other words, the potential has 
flat directions only if the supersymmetric extrema form a valley. If the supersymmetric 
solution represents a discrete set of points, each of them represents an isolated {\it minimum} of the potential. 

Of course, we do not live exactly in Minkowski space, and supersymmetry in our world is broken. In the  simplest KKLT models the gravitino mass $\bar {\it m}$ was of the same order of magnitude as the mass of the volume modulus \cite{Kallosh:2004yh}. The main advantage of the modified version of this mechanism, which was  proposed in \cite{Kallosh:2004yh} and which we are going to generalize below, is that the gravitino mass $\bar {\it m}$ in this scenario  can be many orders of magnitude smaller than the masses of all other fields in Eqs. (\ref{holom}), (\ref{mix}).  For example, all parameters of the modified KKLT construction may have GUT scale or stringy scale, whereas the gravitino mass may be 15 orders of magnitude smaller. In this case our Eqs. (\ref{holom}), (\ref{mix}) imply that a small supersymmetry breaking should not affect the  positive definiteness of the moduli mass matrix. This is the main idea of our work.

\section{Supersymmetric 4d Minkowski vacua of IIB string theory.}

Now let us discuss how one may find the supersymmetric extrema of the potential with respect to all moduli.

Consider a toy model superpotential for moduli stabilization where in addition to the axion-dilaton 
field $\tau$ there are some complex structure fields $x_{i}$ and a total volume $\rho$,
\begin{equation}
  W(\tau, x, \rho)= W_{\rm flux}(\tau, x) + W_{\rm np}(\rho) \ ,
\label{super}
\end{equation}
where, as in Eq. (\ref{flux}), 
\begin{equation}
  W_{\rm flux} = \int G_3\wedge \Omega= f\cdot \Pi(x) -\tau h\cdot \Pi(x) \ .
\label{super2}
\end{equation}
Following \cite{Kallosh:2004yh}, we add to it the racetrack superpotential 
$
W_{\rm np}= Ae^{-a\rho}+ Be^{-b\rho}$
with $a = {2\pi\over N}$, $b = {2\pi\over M}$, where $N$ and $M$ are some integers \cite{racetrack}. The total superpotential is given by
\be
W = f\cdot \Pi(x) -\tau h\cdot \Pi(x)  + Ae^{-a\rho}+ Be^{-b\rho} \ .
\label{adssup}
\ee
 Assume that the potential has a supersymmetric Minkowski minimum at $\rho_{0}$, $\tau_{0}$, and $x_{0}$. We therefore have that
at this minimum 
\be
W=0 \ , \qquad \partial_{\rho}W= \partial_{i}W =\partial_{\tau} W =0 \ .
\label{susy1} \ee

This set of equations resembles the ones obtained in \cite{Giryavets:2004zr} for flux vacua of type IIB 
string theory, but there is an important difference. The models considered in \cite{Giryavets:2004zr} 
did not include the volume modulus, only $\tau$ and $ x$, which can be  stabilized by fluxes. Furthermore 
they considered two types of models, ones with $W_{\rm flux}(\tau, x)$ vanishing at the critical point, 
$W_{ 0}\equiv  W_{\rm flux}(\tau_{0}, x_{0})=0$, and  others where $W_{0}\neq 0$. Correspondingly, there 
were two possible sets of equations:
\be
W_{ 0}=0\ , \qquad \partial_{i}W_{\rm flux} =0\ ,  \qquad \partial_{\tau}W_{\rm flux} =0\ ,
\ee
or
\be W_{ 0}\neq 0\ , \qquad D_{i}W_{\rm flux} =0\ ,  \qquad D_{\tau}W_{\rm flux} =0 \ .
\ee

Meanwhile, in our case the total superpotential  $W(\tau, x, \rho)= W_{\rm flux}(\tau, x) + W_{\rm np}(\rho)$ 
vanishes at the supersymmetric critical point, and covariant derivatives coincide with simple derivatives 
even for $W_{ 0}\equiv W_{\rm flux}(\tau_{0}, x_{0})\not = 0$:
 \be
W_{ 0}\neq 0\ , \qquad \partial_{i}W_{\rm flux} =0\ ,  \qquad \partial_{\tau}W_{\rm flux} =0 \ .
\ee

In the minimum with respect to $x^i$ and $\tau$, the   superpotential  $W_{\rm flux}(\tau, x) + W_{\rm np}(\rho)$ 
acquires the familiar racetrack form  \cite{racetrack}
 \be
W = W_0 + Ae^{-a\rho}+ Be^{-b\rho} \ .
\label{adssup11}
\ee
Equation $\partial_{\rho}W = 0$ in this theory has a following solution \cite{Kallosh:2004yh}:
\be
 \rho_{0}= {1\over a-b}\ln \left |{a\,A\over b\,B}\right|\, .
\label{sigmacr} \ee
 The remaining equations look as follows:
\begin{equation}
\partial_{\tau} W =0 \qquad \Rightarrow \qquad  h\cdot \Pi(x_0) =0 \ .
\label{x0}
\end{equation} 
Therefore
\begin{equation}
 W_{0} \equiv W_{\rm flux}(\tau_{0}, x_{0}) = f\cdot \Pi(x_0) \ .
\label{W0}
\end{equation}
The next set of equations is
\begin{equation}
\partial_{i}W =0 \qquad \Rightarrow \qquad  f\cdot \Pi_{\, , i}(x_0) -\tau_0\, h\cdot \Pi_{\, , i}(x_0) =0 
\label{i}
\end{equation}
which defines the axion-dilaton values at the minimum,
\begin{equation}
\tau_0= {f\cdot \Pi_{\, , i}(x_0)\over h\cdot \Pi_{\, , i}(x_0)} \ .
\label{tau0}
\end{equation} 
Finally, the requirement $W(\tau_{0}, x_{0}, \rho_{0})=0$ leads to a condition
  \be W_0=  -A e^{-a\rho_{0}}- Be^{-b\rho_{0}} =-A \left |{a\,A\over
b\,B}\right|^{a\over b-a} -B \left |{a\,A\over b\,B}\right| ^{b\over b-a}\ .  \ee

One way to solve this set of equations is to find the solutions for $x_{0}$ and $\tau_{0}$ using Eqs. (\ref{x0}), (\ref{tau0}), then calculate $W_{0}$ using Eq. (\ref{W0}), and finally find a set of parameters $A$, $B$, $a$ and $b$ such that $W_0= -A \left |{a\,A\over
b\,B}\right|^{a\over b-a} -B \left |{a\,A\over b\,B}\right| ^{b\over b-a}$.

\section{An example of the solution}\label{example}

To be sure that the mechanism proposed above can actually work, we will construct a particular 
example using the results of Refs. \cite{Kallosh:2004yh,Giryavets:2004zr}. We will consider the model A 
of Ref.  \cite{Giryavets:2004zr} with one complex structure modulus $x_1$, which, following notation
in \cite{Giryavets:2004zr} we will denote by $\psi$. We will solve our equations expanding them 
in series in $\psi^{2}$, which is supposed to be small near the extremum.

According to \cite{Giryavets:2004zr}, at small $\psi$, the period vector in their model can be expanded as 
\be{\Pi_i  =
c_0 p_0 + c_2 p_2\, \psi^2 + c_4 p_4\, \psi^4 + \cdots}\ee
Here
the vectors $p_k$ are given by \be{p_k = ~m_i \cdot \tilde{p}_k}, \quad m_i =\pmatrix{-{1 \over 2} & -{1 \over 2} & {1 \over 2}
& {1 \over 2}\cr 0 & 0 & -1 & 0\cr -1 & 0 & 3 & 2 \cr 0 & 1 & -1 &
0\cr}\ee
with
\be{\tilde{p}_k^T ~= (\alpha^{2(k+1)},
\alpha^{(k+1)}, 1, \alpha^{7(k+1)}), \quad \alpha = e^{i{\pi\over 4}}}\ee and 
\begin{eqnarray}
&& c_0 = (2\pi i)^3 {\sqrt{\pi} \over 2 \Gamma^4(7/8)}
{\exp({7 \pi i\over 8
          })\over \sin({\pi \over 8})},   \nonumber \\
&&c_2 = -(2\pi i)^3 {2\sqrt{\pi} \over  \Gamma^4(5/8)} {\exp({5 \pi i\over 8
          })\over \sin({3\pi \over 8})}, \nonumber \\
&& c_4 = (2\pi i)^3 {4\sqrt{\pi} \over  \Gamma^4(3/8)} {\exp({3 \pi i \over 8
          })\over \sin({5\pi \over 8})}~.
          \end{eqnarray}
The solutions we will present here will be valid for small $\vert \psi \vert$, 
where the $\psi$ modulus will be stabilized in a self-consistent approximation.

We first find $\tau_{0}$, $\psi_{0}$ and $W_{0}$ using Eqs. (\ref{tau0}), (\ref{x0}) and (\ref{W0}) 
for a particular set of parameters used in Ref. \cite{Giryavets:2004zr} namely: \ $f = \{-28, 24, 7, -24\}$, \, $h = \{-34, 41, 12, -17\}$. 
Note that it was shown in \cite{Giryavets:2004zr} that, at the lowest order in $\psi^2$, the moduli space point where
$\partial_{i}W_{\rm flux} =0$ and $\partial_{\tau}W_{\rm flux} =0$ one finds $W_{0} = 0$.
If, however, one solves these equations more precisely, at second order in $\psi^{2}$, one finds a small but nonzero value of $W_{0}$.
 
In \cite{Giryavets:2004zr} it was necessary to solve a system of equations $D_{\psi}W_{\rm flux} =0$, $D_{\tau}W_{\rm flux} =0$,
 each of which depended on $W_{0}$.  However, as we already discussed in the previous section, in our model one should  
solve a simpler set of equations $\partial_{\psi}W_{\rm flux} =0$,  $\partial_{\tau}W_{\rm flux} =0$. 
After finding  $\tau_{0}$ and $\psi_{0}$, one can calculate $W_{0}$ using Eq. (\ref{W0}).

Our investigation gives the following results for $\tau_{0}$, $\psi_{0}$ and $W_{0}$:
 \begin{equation}
\psi_0 \approx 0.088 -0.036\, i,\quad \tau_{0} \approx 1.4\, i, \quad  W_{0}  \approx 0.0766  - 0.1849\, i, \quad |W_{0}| \approx 0.2 \ .
\label{dilcomplmod}
\end{equation} 

The only remaining problem is to find parameters $A$, $B$, $a$ and $b$ such that $W_0= -A \left |{a\,A\over
b\,B}\right|^{a\over b-a} -B \left |{a\,A\over b\,B}\right| ^{b\over b-a}$. Obviously, there are many sets of parameters satisfying this condition. One particular set is
 \begin{equation}
a ={2\pi\over 200},\quad b ={2\pi\over 100}, \quad  A  = - \beta, \quad B = 10 \beta,  \label{volparam}
\end{equation} 
where $\beta = 3.06 - 7.39 \, i$. Fig. 1 shows the shape of the potential for the volume modulus for this set of parameters.

We plot the slice of the total potential 
\be
V(\tau, \bar\tau; \psi, \bar\psi; \rho,\bar \rho)= e^{K(\tau, \bar\tau; \psi, \bar\psi)}e^{K(\rho, \bar\rho)}\left(  |D_\tau W|^2+ |D_\psi W|^2+ |D_\rho W|^2-3|W|^2\right) 
\ee
at the critical values of $\tau_0$ and $\psi_0$ given in eq. (\ref{dilcomplmod})

 \begin{figure}[h!]
\centering\leavevmode\epsfysize=6 cm \epsfbox{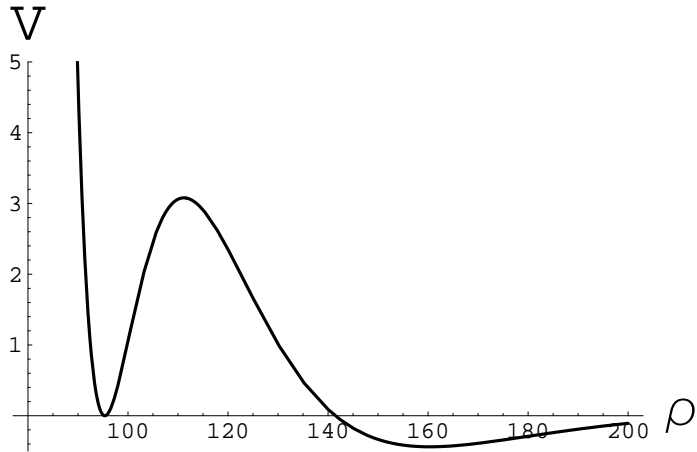}

\

\caption{Potential of the volume modulus for $a ={2\pi\over 200}$, $b ={2\pi\over 100}$, $A  = - \beta$, $B = 10 \beta,$
where $\beta = 3.06 - 7.39 \, i$. The potential is shown in units of $10^{-15}M^{4}_{p}$.}\label{fig:Fig1}
\end{figure}
The scale of the potential is defined by the value of $e^{K(\tau, \bar\tau; \psi, \bar\psi)}e^{K(\rho, \bar\rho)} W_0$. It is interesting that 
 in our example $e^{K(\tau, \bar\tau; \psi, \bar\psi)}$ and $e^{K(\rho, \bar\rho)}$ are of the same order. Therefore, when considering a truncated system where only $\rho$ modulus is present, one has to use for effective $W_0$ the value of $e^{K} |W_{flux}|^2(f, h, \tau_0, \bar\tau_0; \psi_0, \bar\psi_0)$ as proposed in \cite{Giryavets:2004zr}. Indeed in our case $e^{K(\tau_0, \bar\tau_0; \psi_0, \bar\psi_0)}\sim 10^{-7}$ which is quite striking: it indicates that the  K\"ahler potential due to complex structure moduli   may become a dramatic source of a change of the scale of the potentials in various models.

The model presented above preserves all good features of the KKLT construction and has all moduli stabilized due to supersymmetry of the Minkowski vacuum state at $\rho_{0} \approx 95$.

\section{Conclusions.}

The model described in this paper is somewhat more complicated than the original KKLT construction, 
but it has several advantages. First of all, the scale of supersymmetry breaking in this model can be 
hierarchically small as compared to stringy scale and to the scale of all moduli masses stabilized by the KKLT mechanism.
This fact ensures vacuum stability, which otherwise would require a detailed study involving the investigation of
convexity of the potential of all moduli, a rather daunting task with an uncertain outcome. Thus, at the very least, 
this model can be useful as a playground for investigation of the regions of the stringy landscape where the vacuum 
stability is protected by supersymmetry in Minkowski space. The smallness of supersymmetry breaking  was previously
 postulated as a technical tool for solving the hierarchy problem and explaining smallness of the Higgs boson mass. 
It is quite intriguing that in string theory this condition can be linked to the requirement of vacuum stability. 

It is equally interesting that the requirement of the hierarchically small supersymmetry breaking allows one to have 
inflation on a very high energy scale. This alleviates the problem of constructing unusual inflationary models with 
an extremely small scale of inflation. This also simplifies a solution of the problem of initial conditions for 
inflation, which is much easier to achieve for the high energy scale inflation.

In this paper we considered a model with the nonperturbative racetrack potential $Ae^{-a\rho}+ Be^{-b\rho}$. 
However, the main idea of our work is more general: One should  study all possible nonperturbative and perturbative
corrections to the superpotential, and then try to find a supersymmetric state where the total superpotential 
(and, consequently, the total potential) vanishes. This problem is technically simpler than the more general 
problem of finding a supersymmetric AdS extremum because in Minkowski vacuum the covariant derivatives 
involving K\"ahler potential and superpotential reduce to the usual derivatives. Once the solution is found, 
it is automatically stable due to supersymmetry, the potential is convex, and this property should be preserved
 after the remaining small uplifting.

\

\leftline{\bf Acknowledgments}

The authors are  grateful to the organizers of the conference ``The Next Chapter in Einstein's Legacy'' and the 
subsequent workshop at the Yukawa Institute, Kyoto, where this work was initiated. We are also grateful to  A. Giryavets, M. Gomez-Reino, K.~Choi, 
M. Dine, S. Kachru, O. Pujolas,  M. Redi, M. Sasaki,  P.K. Tripathy and S.P. Trivedi for useful discussions. J.J. B-P. is supported by the James Arthur Fellowship at NYU. 
The work of R.K and A.L. was supported by NSF grant 0244728 and by Kyoto University. 

\

\end{document}